# Processor-Dependent Malware... and codes[*]

Anthony Desnos[1], Robert Erra[2], and Eric Filiol[1]

[1] ESIEA - $(C+V)^O$, 38 rue des Dr Calmette et Guérin, 53 000 Laval, France
{desnos,filiol}@esiea.fr
[2] ESIEA - SI&S, 9 rue Vsale, 75 005 Paris, France, erra@esiea.fr

**Abstract.** Malware usually target computers according to their operating system. Thus we have Windows malwares, Linux malwares and so on ...In this paper, we consider a different approach and show on a technical basis how easily malware can recognize and target systems selectively, according to the onboard processor chip. This technology is very easy to build since it does not rely on deep analysis of chip logical gates architecture. Floating Point Arithmetic (FPA) looks promising to define a set of tests to identify the processor or, more precisely, a subset of possible processors. We give results for different families of processors: AMD, Intel (Dual Core, Atom), Sparc, Digital Alpha, Cell, Atom ... As a conclusion, we propose two *open problems* that are new, to the authors' knowledge.

## 1 Introduction

From the beginning of malware history (circa 1996), malware are:

- either operating system specific (Windows *.*, Unices, Mac, ...);
- or application specific (e.g. macro viruses);
- or protocol dependent (e.g. *Conficker* versus *Slammer*).

We will use the following and large definition of a malware: it is a *malicious code* like a virus, a worm, a spyware, a Trojan horse... whose aim is to undermine system's confidentiality, intergrity or availability.

At the present time, there are quite no hardware specific malwares, even if some operating system are themselves hardware dependent (e.g. *Symbian* malwares). Recently, GPGPU malware [IPV10] have been proposed but they just exploit the fact that graphic cards are just distinct devices with quite the same features and capability as the system they are connected to. They do not really condition their action on new computing features. GPGPU malware strongly depend on the graphic card type (CUDA or OpenCL enabled).

We propose here to investigate the following critical issue: *is it possible to design malware – or more generally, any program – that operate beyond operating system and application types and varieties*? More precisely, we want:

---

[*] The present paper is the extended version of the work presented at the *i*AWACS'09. Conference

- going beyond operating system and application types/varieties ... ;
- while exploiting hardware specificities.

If such an approach is possible, this would:

- enable far more precise and targeted attacks, at a finer level (surgical strikes) in a large network of heterogeneous machines but with generic malware;
- and represent a significant advantage in a context of cyberwarfare.

The recent case of the *StuxNet* worm shows that targeted attacks towards PLC components are nowadays a major concern in cyberattacks. However, while it can be very difficult to forecast and envisage which kind of applications is likely to be present on the target system (it can be a secret information), the variety in terms of hardware – and especially as far as processors are concerned – is far more reduced due to the very limited number of hardware manufacturers. We propose to consider *processor-dependent malware* and to rely on the onboard processor, which seems a good candidate to design hardware dependent software (a malware is indeed a software).

To design such *dependent processor malware*, we need to identify the processor as precisely as possible. This is possible thanks to a different ways:

- by reversing existing binaries (but this provides a limited information since a given binary can indifferently execute on several processors like Intel x86s or AMDs chips),
- classical intelligence gathering...

There is a large spectrum of possibilities to collect this technical intelligence. But there is a bad news: deriving knowledge about processor internals is tricky and require a lot of work. Instead of analyzing processor logic gates architecture, we propose to work at the higher level: *to exploit mathematical perfection versus processor reality*.

This paper is organized as follows. Section 2 sets up the theoretical background which make hardware-dependent malware possible. Then Section 3 exposes how to exploit processors' mathematical limitations in order to make programs' execution vary according to the processor in place. Section 4 then gives implementation and experimental results before concluding and presenting future work in Section 5.

## 2 Theoretical Background

### 2.1 Starting From a Formal Model of Malware - Notation

We consider the formal model given by Zuo and Zhou in 2004 [ZZ04], Zuo, Zhou and Zhu in 2005 [ZZZ05] and Filiol in 2004 [Fil05].

- Sets $\mathbb{N}$ and $S$ are the set of natural integers and the set of all finite sequences of such integers, respectively.
- Let $s_1, s_2, \ldots, s_n$ be elements from $S$.

- Let $< s_1, s_2, \ldots, s_n >$ describe an injective computable function from $S^n$ to $\mathbb{N}$ whose inverse function is computable as well.
- If we consider a partial computable function $f : \mathbb{N} \to \mathbb{N}$, then $f(s_1, s_2, \ldots, s_n)$ describes $f(< s_1, s_2, \ldots, s_n >)$ in an abridged way.
- This notation extends to any $n$-tuple of integers $i_1, i_2, \ldots, i_n$.

- For a given sequence $p = (i_1, i_2, \ldots, i_k, \ldots, i_n) \in S$, we denote $p[j_k/i_k]$ the sequence $p$ in which the term $i_k$ has been replaced by $j_k$, let say $p[j_k/i_k] = (i_1, i_2, \ldots, j_k, \ldots, i_n)$.
- If the element $i_k$ of sequence $p$ is computed by a computable function $v$ (equivalently compute $p[v(i_k)/i_k]$), let us adopt the equivalent abridged notation $p[v(\underline{i_k})]$ in which the underlined symbol describes the computed element.
- In the general case (compute more than one element at the same time in $p$), we note $p[v_1(\underline{i_{k_1}}), v_2(\underline{i_{k_2}}), \ldots, v_l(\underline{i_{k_l}})]$.

Now that everything to model programs has been given, let us define things at a higher level formally: program, data and operating system.

- We describe by $\phi_P(d, p)$ a function which is computed by a program $P$ in the environment $(d, p)$.
  - $d$ and $p$ are denoting data in the environment (including clock, mass memories and equivalent structures or devices) and programs (including those of the operating system itself) respectively.
  - That environment corresponds in fact to the operating system which has been extended to the activity of one or more users.
- When considering the Gödel coding $e$ for the program $P$, we use the notation $\phi_e(d, p)$. Its definition domain is then denoted by $W_e$ while his image space is denoted $E_e$.

### 2.2 Exploring the Viral Classes

Let us give the general formal definition of computer viruses (most complete case) with the following definition. However this definition can be extended to any other kind (non self-reproducing) malware and more generally to any program, eventually by dropping the self-reproduction properties off.

**Definition 1** *(Non Resident Virus) A total recursive function $v$ is a non resident virus if for every program $i$, we have:*

1. $\phi_{v(i)}(d, p) = \begin{cases} D(d, p), & \text{if } T(d, p) \ (i) \quad \text{(Added Fonctionnality)} \\ \phi_i(d, p[v(\underline{S(p)})]) & \text{if } I(d, p) \ (ii) \quad \text{(Infection)} \\ \phi_i(d, p), & \text{otherwise } (iii) \ \text{(Imitation)} \end{cases}$
2. $T(d, p)$ and $I(d, p)$ are two recursive predicates such that there is no value $< d, p >$ that satisfies them both at the same time. Moreover both functions $D(d, p)$ et $S(p)$ are recursive.
3. The set $\{< d, p >: \neg(T(d, p) \vee I(d, p))\}$ is infinite.

The two predicates $T(d,p)$ and $I(d,p)$ represent the payload and the infection trigger conditions respectively. Whenever $T(d,p)$ is true, the virus executes the payload $D(d,p)$ while whenever $I(d,p)$ is true, the virus selects a target program by means of the selection function $S(p)$ and then infects it. Finally the original program $i$ is executed (host program). For a virus kernel: the set of functions $D(d,p)$ and $S(p)$ with predicates $T(d,p)$ and $I(d,p)$: the virus kernel describes the malware in a univoqual way. This model can be extended to other form of malware (more sophisticated viruses, Trojan... ).

**Polymorphic and Metamorphic Viruses**

**Definition 2** *The pair $(v, v')$ of total recursive functions $v$ and $v'$ is called Polymorphic virus with two forms if for every program $i$ we have*

$$\phi_{v(i)}(d,p) = \begin{cases} D(d,p), & \text{if } T(d,p) \\ \phi_i(d,p[v'(\underline{S(p)})]), & \text{if } I(d,p) \\ \phi_i(d,p), & \text{otherwise} \end{cases}$$

*and*

$$\phi_{v'(i)}(d,p) = \begin{cases} D(d,p), & \text{if } T(d,p) \\ \phi_i(d,p[v(\underline{S(p)})]), & \text{if } I(d,p) \\ \phi_i(d,p), & \text{otherwise} \end{cases}$$

Whenever predicate $I(d,p)$ is true the virus selects a target program by means of $S(p)$, infects it then transfers control back to the host program $x$. $S(p)$ is performing the code mutation as well.

**Definition 3** *Let $v$ and $v'$ be two different total recursive functions. The pair $(v, v')$ is called metamorphic virus if for every program $i$, then the pair $(v, v')$ satisfies:*

$$\phi_{v(i)}(d,p) = \begin{cases} D(d,p), & \text{if } T(d,p) \\ \phi_i(d,p[v'(\underline{S(p)})]), & \text{if } I(d,p) \\ \phi_i(d,p), & \text{otherwise} \end{cases}$$

*et*

$$\phi_{v'(i)}(d,p) = \begin{cases} D'(d,p), & \text{if } T'(d,p) \\ \phi_i(d,p[v(\underline{S(p)})]), & \text{if } I'(d,p) \\ \phi_i(d,p), & \text{otherwise} \end{cases}$$

*where $T(d,p)$ – respectively $I(d,p)$, $D(d,p)$, $S(p)$ – is different from $T'(d,p)$ – respectively $I'(d,p)$, $D'(d,p)$, $S'(p)$.*

Metamorphic viruses are similar to polymorhic viruses except that selection functions $S(p)$ and $S'(p)$ are different. The kernel of metamorphic forms are totally different.

**Stealth Viruses**

**Definition 4** *The pair $(v, sys)$ made of a total recursive function $v$ and a system call sys (a recursive function as well) is a stealth virus with respect to the system call sys, if there exists a recursive function $h$ such that for every program $i$ we have:*

$$\phi_{v(i)}(d,p) = \begin{cases} D(d,p), & \text{if } T(d,p) \\ \phi_i(d, p[v(\underline{S(p)}), h(\underline{sys})]) & \text{si } I(d,p) \\ \phi_i(d,p), & \text{otherwise} \end{cases}$$

*et*

$$\phi_{h(sys)}(i) = \begin{cases} \phi_{sys}(y), & \text{if } x = v(y) \\ \phi_{sys}(i), & \text{otherwise} \end{cases}$$

Let us point out that *stealth* is a relative concept (with respect to a given set of system calls).

### 2.3 Practical Utility of the Formal Model: What Does the Model Show Us

We must identify and use a feature that will make a virus (in the general case, a malware) operate whether a given type of processor chip is present or not. In the previous formal definition, whatever may be the class of virus, the obvious candidates for usable features are predicates $T(d,p)$ and $I(d,p)$ (payload and infection trigger conditions respectively). In the optimal case, we are interested in considering two different features to control and manage payload triggering and infection control separately and independently. So:

– Code mutation and stealth can also be managed with respect to specific processors in the same way.
  • As an example a malware will enforce Hardware Virtual Machine-based rootkit techniques whenever present.
  • Code mutation (e.g metamorphism) will be activated only if a suitable processor instruction set is available.
– This approach, yet formal, gives a powerful insight of how design processor-dependent malware.
– This enables to reduce the problem of side effects significantly, that may betray the activity of a malware.

## 3 Exploiting Mathematical Processor Limitations

In order to use processor to discriminate programs' action and execution, we are going to exploit the fact that first there is a huge difference between the mathematical reality and their implementation in computing systems and second that that difference is managed in various ways according to the processor brand, model and type.

## 3.1 Mathematical perfection versus Processor Reality

Let us begin with a very classical example: the algorithm given in Table 1. We can ask: *what does this code (really) compute?*

---

**Algorithm** 1 : The $\sqrt{\phantom{x}}$ problem
   **Input**: — a real $A$;
   **Output**: — a boolean $B$
   **Begin**:
      $B = \sqrt{A} * \sqrt{A}$;
      **Return**[A==B];
   **End.**

---

**Table 1.** The Square-root problem

Well, let us suppose we choose $A = 2.0$ as input for this *Square-root* algorithm, we then have two possible answers, that are quite *opposite*:

1. *Mathematically*: **True** is returned;
2. *Practically*: **False** is returned!

Let us now explain why we have this different output. This come from the fact that processors:

- have an increasing (architecture) complexity and size,
- have bugs, known and unknown (not published),
- use floating point arithmetic,
- use generally "secret" algorithms for usual arithmetic functions like $1/x, \sqrt{x}$, $1/\sqrt{x}$ ... that can be computed:
  1. at the *hardware* level;
  2. and/or at the *software* level.

As an example of a "secret algorithm", let us cite the famous Pentium Bugs *case* in 1994: Intel has never published neither the *correct* algorithm nor its bugged version used for the division but some researchers have tried reverse engineering techniques to understand which algorithm was programmed actually (for instance, the reader will refer to the (beautiful) paper [CMMP95]).

Let us now consider the following problem: *can we define a set of (simple) tests to know on which processor we are?* As a practical example: *is it possible to know whether we are on a mobile phone or on a computer?*

The Intel Assembly Language instruction **CPUID** can be used both on Intel and AMD processors, but it has at least two severe drawbacks:

- it is easy to "find" it whenever scanning the file (malware detection issue);
- some other processors cannot recognize and process this instruction.

### 3.2 Processor Bugs

Known or unknown bugs are good candidates to design such a set of tests and hence to discriminate processors:

- as an instance of such bug/test, it is easy determine whether we use a 1994 bugged Pentium or not: just use the numerical value that makes appear the *Pentium Division Bug*;
- but a lot of bugs will *freeze* the computer only (this can be used for processor-dependent denial of service [DoS] however);
- and it is not so simple to find a list of bugs, even if there are supposed to be "known".

So in this paper, we will not use bugs *unless* they involve a floating point arithmetic operator. However it is worth keeping in mind that the knowledge of some bugs (by the manufacturer, a Nation State...) can be efficiently used to target processors specifically and hence it represents a critical knowledge not to say a strategic one. Worse, hiding such bugs or managing floating arithmetics in a very specific way is more than interesting.

More generally let us consider some differences that exist event within the same type of processors but produced in two different versions: a national and an "export" version. As an example, we can consider the POPCOUNT function which compute the Hamming weight of an integer (the number of 1s in its binary form). Since it is a critical function in the context of cryptanalysis, the national version of a few processors have this function implemented in hardware while the export version just emulate it at the software level. Consequently a good way to discriminate national version from export version consists in computing Hamming weight a large number of times and then to record the computation time: it will be significantly higher for the export version which hence can be specifically targeted by a malware attack.

### 3.3 Using Floatinf Point Arithmetics: The IEEE P754 Standard

The IEEE P754 standard [Ove01] has been approved as a norm by IEEE ANSI in 1985. A lot of processors follow and comply to it but some processors do not. As an example, let us mention the CRAY 1 or DEC VAX 780. Moreover, not all microcontrollers follow this standard either.

This norm does not impose the algorithms to compute usual functions lke $1/x, \sqrt{x}, 1/\sqrt{x}$ or $e^x$. It just gives a specification for the four basic operations: addition, substraction, multiplication and division. So, for all other functions, there is very likely to exist differences as far as their implementation as algorithms are concerned. But we have to find them!

For 32-bit, environments, we have (see Table 2):

- 1 bit for the sign;
- 23 bits for the mantissa;
- 8 bits for the exponent (integer).

| sign(x) | mantissa(x) | exponent(x) |
|---------|-------------|-------------|
| 1 bit   | 23 bits     | 8 bits      |

**Table 2.** Structure of 32-bit float "numbers" in the IEEE P754 Standard

The floating point arithmetic has a lot of curiosities, let us see some of them. One can find in [DM97,KM83] the following questions due to Rump:

- Evaluate the expression

$$F(X, Y) = \frac{(1682XY^4 + 3X^3 + 29XY^2 - 2X^5 + 832)}{107751}$$

with $X = 192119201$ and $Y = 35675640$. The "exact" result is $1783$ but numerically we can have a very different value like $-7.18056\,10^{20}$ (on a 32-bit IEEE P754 compliant processor).

- Evaluate the expression

$$P(X) = 8118X^4 - 11482X^3 + X^2 + 5741X - 2030$$

with $X = 1/\sqrt{2}$ and $X = 0.707$. The "exact" result is $0$ but numerically we can have a very different value like $-2.74822\,10^{-8}$ (on a 32-bit IEEE P754 compliant processor).

Let us recall that the numerical value of an algebraic expression depends (generally) on the compiler because a non basic numerical expression result depends strongly on the order of the intermediate computations.

## 4 Implementation and Experimental Results

Now we have defined and illustrated the core principle of our approach, let us consider a few implementations we have considered as well as the corresponding results.

### 4.1 The Gentleman Code or How to Compute the Word Length

If we want to know on which processor we are working, we need to find, before anything else, two critical information:

1. the first thing is to find the *base* value used to represent numbers;
2. the second is the *word length*, *i.e.* the number of bits the processor is used to work (with floating point numbers for example).

For the base value, it is easy to conjecture that the base is 2, at least for modern processors. As far as the the word length is concerned, we have not found any numerical algorithm that is able to answer this question but we have found something very close. The algorithm given in Table 3 called the *Gentleman Code* [GM74,Mul89] is surprisingly very interesting for both problems. First we can again ask: *what does this code (really) compute?* Well, again, we have two possible answers:

```
Algorithm 2 :   The Gentleman Code
   Input: — A=1.0 ; B=1.0;
   Output: — A, B
   Begin:
      A=1.0;
      B=1.0;
      While ((A+1.0)-A)-1.0==0 ;
         A=2*A;
      While ((A+B)-A)-B==0 ;
         B=B+1.0;
      Return[A,B];
   End.
```

**Table 3.** The Gentleman code

1. *Mathematically*: the two loops are theoretically *infinite loops* so they are looping forever;
2. *Practically* (see [Mul89]):
   - $\log_2(A)$ gives the number of bits used by the mantissa of floating point numbers;
   - $B$ is the base used by the floating point arithmetic of the environment (generally it is equal to 2).

Both values are of course *processor-dependent* constants. So, with a small program, which has a polynomial time complexity, we can compute the number of bits used to represent the *mantissa* of any floating point number and so, we can deduce the word length.

### 4.2  Some Basic (but Yet Too) Simple Tests

Let us give a first list of tests we have tried (Table 4). So these tests are interesting but not completely useful, this shows that we can simply know whether the processor follows the IEEE P754 arithmetic norm or not. For these simple expression, all processors that are IEEE P754 compliant will give the same answers. Hence, this is not enough.

### 4.3  Some Less Basic Tests

With the following constant definitions in our test program in C, we obtain the results given in Tables 5 and 6.

- #define Pi1 3.141592653
- #define Pi2 3.141592653589
- #define Pi3 3.141592653589793

| Processor | Tests | | | |
|---|---|---|---|---|
| | 1.2-0.8 == 0.4 | 0.1+0.1 == 0.2 | 0.1+0.1+0.1 == 0.3 | 0.1+...0.1 == 1.0 |
| VAX 750 | Yes | Yes | No | No |
| AMD 32 | No | Yes | No | No |
| AMD 64 | No | Yes | No | No |
| ATOM | No | Yes | No | No |
| INTEL DC | No | Yes | No | No |
| MIPS 12000 | No | Yes | No | No |
| dsPIC33FJ21 | No | Yes | Yes | No |
| IPHONE 3G | No | Yes | No | No |

**Table 4.** A few easy computations

– #define Pi4 3.1415926535897932385

These results are more interesting, especially those in the third column (the numerical computation of $\sin(10^{37}\pi_1)$) in Table 5: a simple computation gives four subclasses of the set of processors (emphasized by a double horizontal lign between the subclasses).

| Processor | $\sin(10^{10}\pi_1)$ | $\sin(10^{17}\pi_1)$ | $\sin(10^{37}\pi_1)$ | $\sin(10^{17}\pi_1) == \sin(10^{17}\pi_2)$ |
|---|---|---|---|---|
| IPHONE 3G | 0.375... | 0.423... | -0.837... | No |
| AMD 32 | 0.375... | 0.424... | -0.837... | No |
| AMD 64 | 0.375.. | 0.424.. | 0.837... | No |
| ATOM | 0.375.. | 0.423.. | -0.832.. | No |
| INTEL DC | 0.375... | 0.423... | -0.832... | No |
| MIPS 12000 | 0.375... | 0.423... | -0.832... | No |
| dsPIC33 | *0.81...* | *0.62...* | *-0.44...* | *Yes* |

**Table 5.** Computation of $\sin(10^{10}\pi)$ for various numerical values of the constant $\pi$

| Processor | $\sin(10^{37}\pi_1)$ | $\sin(10^{37}\pi_2)$ | $\sin(10^{37}\pi_3)$ | $\sin(10^{37}\pi_4)$ |
|---|---|---|---|---|
| IPHONE 3G | 47257756 | 9d94ef4d | 99f9067 | 99f9067 |
| AMD 64 | af545000 | af545000 | af545000 | af545000 |
| ATOM | 47257756 | 9d94ef4d | 99f9067 | 99f9067 |
| INTEL DC | 47257756 | 9d94ef4d | 99f9067 | 99f9067 |
| MIPS 12000 | 47257756 | 9d94ef4d | 99f9067 | 99f9067 |
| dsPIC33 | bee5 | bee5 | bee5 | bee5 |

**Table 6.** $\sin(10^{37}\pi)$ in hex for various numerical values of the constant $\pi$

### 4.4 Do Not Forget the *Influence* of the Compiler

Let us give a last example. We want to compute the generalized sum

$$s(N) := \sum_{i=1}^{N} 10^N \qquad (1)$$

The "exact" value is of course $N * 10^N$, but let us have a look at the Table 7 to see some values we can have when computing $s(N) - N * 10^N$. However we have to point out that the results of the Table 7) heavily depend of course of the processor but *also* of the compiler used, of the options used and so on ...

| N | 10 | 21 | 22 | 25 | 30 | 100 |
|---|---|---|---|---|---|---|
| $s - N * 10^N$ | 0.0 | 0.0 | $-8.05\ 10^8$ | $-6.71\ 10^7$ | $-4.50\ 10^{15}$ | $4.97\ 10^{86}$ |

**Table 7.** Computation of $s(N) - \sum_{i=1}^{N} 10^N$ for different values of $N$

More work has to be done to better understand these aspects. Nonetheless, we have here new insights on how design more specific attacks when considering the processor type AND the compiler version/type as the same time.

## 5 Conclusion and Future Work: Two interesting Open Problems

Floating Point Arithmetic (FPA) looks promising to define a set of tests enabling to identify the processor or, more precisely, a subset of possible processors. We intend to propose very soon the Proc_Scope Tool: a sotfware tool. Proc_Scope uses carefully chosen *numerical expressions* that give information on the processor type.

More results will be published very soon. We propose now two *open problems* that, to the authors' knowledge, are new.

The first open problem we propose is the one discussed in this work: *can we find an numerical algorithm, with a linear complexity in time and space and compute a floating point expression, that can help to distinguish a given processor precisely?* Beyond the examples we have proposed here, a promising algorithm could be based on a variant of the famous *logistic equation*, thoroughly studied in the chaos theory, which is defined by:

$$x_{n+1} = r\, x_n\, (1 - x_n) \qquad (2)$$

with $r \in [0, 4]$.

The sequence defined by Equation 2, for a chosen and fixed $x_0$, can exhibit very different behaviors:

- a *periodic* behavior for example for values of r less than 3.0;
- or a *chaotic* behavior for values of $r$ slightly larger than 3.57.

See [ASY96,Bar88,Dev86,PJS92] for a detailed study of the properties of these sequences.

Finally, we propose another new problem: *find processor-dependent hash functions*. Generally, hash functions are defined as independent from the processor. But, in some cases, one can desire to get rid of this view. We propose in fact to take the *opposite idea*: we want a hash function that heavily depends of the processor used to compute it. For example, it can be interesting to design a specific hash function for a *smartphone* or a specific processor. The best way to design such a hash function seems to use the properties of the floating point arithmetic operators of the processor; more specifically some of the arithmetic functions implemented on the processor. The second open problem we propose is then: *can we define, for a specific processor, hash functions that use the floating point arithmetic of the concerned processor that respect the classical requirements for such functions?*

## Acknowledgment

The authors thank Olivier TUCHON for his careful reading of this work and his suggestions which have greatly help to improve this paper.